\documentclass[pra,aps,twocolumn]{revtex4}
\usepackage{graphicx}
\usepackage{amssymb}
\usepackage{latexsym}

\def\be{\begin{equation}}
\def\ee{\end{equation}}
\def\bea{\begin{eqnarray}}
\def\eea{\end{eqnarray}}

\def\d{{\,\rm d}}

\def\n{{\bf n}}
\def\m{{\bf m}}

\def\p{{\bf p}}

\def\sgn{\,{\rm sgn\,}}

\def\h2m{\frac{\hbar^2}{2m}}
\def\p0{{P_{\beta H^0_N}}}

\begin{document}

\title{\large\bf Structure factor and dynamics of the helix-coil transition}

\author{Herv\'e Kunz$\,^1$, Roberto Livi$\,^{2}$,
         and Andr\'as S\"ut\H o$\,^{3}$\\[-2mm]$ $}
\address{
        $^1$
        Institute of Theoretical Physics, Ecole Polytechnique F\'ed\'erale de
        Lausanne, CH-1015 Lausanne, Switzerland\\
        $^2$
        Dipartimento di Fisica, Universit\'a di Firenze,
        INFM and INFN, Firenze, via Sansone 1  I-50019 Sesto Fiorentino, Italy\\
        $^3$
        Research Institute for Solid State Physics and Optics,
        Hungarian Academy of Sciences, P. O. Box 49, H-1525
        Budapest 114, Hungary\\
        }
\date{\today}
\thispagestyle{empty}
\begin{abstract}
\noindent

Thermodynamical properties of the helix-coil transition were successfully
described in the past by the model of Lifson, Poland and Sheraga. Here we
compute the corresponding structure factor and show that it possesses a
universal scaling behavior near the transition point, even when the transition
is of first order. Moreover, we introduce a dynamical version of this model,
that we solve numerically. A Langevin equation is also proposed to describe
the dynamics of the density of hydrogen bonds. Analytical solution of this
equation shows dynamical scaling near the critical temperature and predicts
a gelation phenomenon above the critical temperature.
In the case when comparison of the two dynamical approaches
is possible, the predictions of our phenomenological
theory agree with the results of the Monte Carlo
simulations.

\vspace{2mm}

\vspace{2mm}
\noindent

\end{abstract}
\maketitle

\section{Introduction}

Ever since the discovery of the helicoidal structure of DNA under the usual
conditions, the phenomenon of its transformation to a coil under heating has
been studied and described as an equilibrium phase transition.
Basic works by Lifson and subsequently by Poland and Sheraga led to a model
(LPS) considered today standard \cite{L,PS}.
The model is one-dimensional and
shows a phase transition only because the interactions are long-ranged. A more
realistic description of the two strands moving and interacting in
three dimensional space has also been proposed and studied numerically
\cite{Caruso,Carlon, Baiesi,BaKa} .
A model of LPS--type was shown to be able to consistently fit
all relevant simulation results \cite{Schafer}.
More recently, some authors have proposed a new mechanism which yields an
abrupt first--order phase transition \cite{Peliti, Blossey},
at variance with the previous theoretical considerations in favor of a continuous
one.
On the experimental side the situation is still far from conclusive.
Previous measurements indicated the presence of a sharp jump in the
fraction of bound base pairs \cite{Wartell, Blake}. More recently,
the results reported in \cite{Zocchi} have been interpreted as an indication of a
weakly continuous phase transition.

It seems, however, that little is known about the dynamics of
this denaturation transition. We can mention the contributions based
on a mechanical approach \cite{Bishop2,Theodora} and on a Langevin
description of the two DNA--strands as polymers in continuous space
\cite{Nelson}.

In this paper we investigate the
original LPS model while considering the exponent $\alpha$ as
a free parameter. For what concerns the equilibrium properties
we compute the structure factor and find that near the critical
point it exhibits universal scaling behavior, which depends only on
$\alpha$. Moreover, we also look at the transition from a dynamical
point of view. We do it in two ways: first we describe this dynamics
by a master equation such that the equilibrium state is chosen to be
the one of the LPS model. Secondly, we introduce a Langevin type
equation for the order parameter which is the density
of hydrogen bonds. The analysis of the solution of this equation reveals
a critical slowing-down, the relaxation time
diverging at the critical point with a power law still depending
on $\alpha$. In the absence of noise, we find that above the critical
temperature a {\it gelation phenomenon} occurs, namely the density
of H-bonds approaches zero in a finite time. With the addition of noise
we obtain also the analytic expression for the power--spectrum of the
density of H-bonds. Finally, we compare the two descriptions and find
that the dynamics of the order parameter derived from the master
equation and from the Langevin equations essentially agree.

The following three sections are devoted to a study of equilibrium
properties. In Section II we summarize the basics of the LPS model.
Formulas for the free energy and the pressure are given here, as well
as scaling formulas for the density in the vicinity of the phase transition.
Section III contains the computation of the one- and two-point correlation
functions for finite systems and in the thermodynamic limit. Our results for
the structure factor are presented in Section IV. Results on the dynamics
are collected in Sections V and VI. In Section V we describe the microscopic
dynamics realized numerically in this work. The equation for the order parameter
and its analysis both with and without noise is given in Section VI, where we also
compare the results of the two dynamical approaches.
The paper ends with a Summary.

\section{The LPS model}\label{LPS}

Let 0 and 1 denote respectively a broken or existing H-bond at a given site.
Then, a configuration of the double-strand of DNA of length $N$ is represented
by a sequence $\n=\{n_x\}_{x=1}^N$ where
$n_x=0$ or 1. In $\n$ there are alternating maximal connected subsequences
(intervals) of 1 and 0, respectively.
A weight $a(j)>0$ or $b(j)>0$ is associated to an interval of length $j$ of 1 or
0, respectively. If the lengths of the intervals of $\n$ are $l_i$ and $m_j$ then
the weight of $\n$ is
\be\label{Gibbs}
e^{-V(\n)}=\prod_i a(l_i)\prod_j b(m_j)\ .
\ee
As usual, thermal equilibrium can be discussed in the canonical or in the
grand-canonical ensemble.
In the first case the description is restricted to configurations with the
total number of H-bonds, $n=\sum_xn_x$, fixed. In the second case all configurations
compatible with the boundary condition
are considered but an activity $e^{\beta\mu}$ is introduced.
The chemical potential $\mu$ is the energy
needed to break a H-bond. The canonical partition function is
\be
Q_{N,n}=\sum_{\n:\sum_x n_x=n}   e^{-V(\n)}
\ee
and the grand-canonical partition function is
\be\label{grand-can}
Q_N(\beta\mu)=\sum_{n} Q_{N,n}e^{n\beta\mu}=
\sum_{\n}   e^{-U(\n)}
\ee
where $U(\n)=V(\n)-\beta\mu\sum_x n_x$. Alternately, $e^{-U(\n)}$ is obtained
from $e^{-V(\n)}$ by replacing each factor $a(l)$ by $a(l)e^{l\beta\mu}$.
The free energy $f$ and the pressure $p$ are respectively given by
\be\label{f}
\beta f=-\lim_{\begin{array}{cc}
N\to\infty\\n/N\to\rho\end{array}
}\frac{1}{N}\ln Q_{N,n}
\ee
and
\be\label{p}
\beta p=\lim_{N\to\infty}\frac{1}{N}\ln Q_N(\beta\mu).
\ee
The temperature
does not appear explicitly in the canonical partition function, therefore $\beta f$
is independent of $\beta$ and $\beta p$ depends on it through
$\epsilon=\beta\mu$.
The free energy and the pressure
are Legendre transforms of each other,
\be\label{L1}
\beta p(\epsilon)=\sup_{\rho}\{\epsilon\rho-\beta f(\rho)\}\equiv
\epsilon\rho_\epsilon-\beta f(\rho_\epsilon)
\ee
and
\be\label{L2}
\beta f(\rho)=\sup_{\epsilon}\{\epsilon\rho-\beta p(\epsilon)\}
\equiv\epsilon_\rho\rho-\beta p(\epsilon_\rho)
\ .
\ee
Here $\rho$ is the density of 1 (H-bonds), and
$\rho_\epsilon$ and
$\epsilon_\rho$ are the respective values at which the suprema
are attained.

The partition functions depend on the boundary conditions imposed on the system.
Typical choices are the free and the periodic conditions, or fixed 0 or 1 on the
boundary sites. For example, with the choice $0\ldots 1$ or $1\ldots 0$,
\be\label{can}
Q_{N,n}=\sum_{k\geq1} A_k(n)B_k(N-n)
\ee
where
\be
A_k(n)=\sum_{l_1\geq 1}\cdots\sum_{l_k\geq 1}\prod_{i=1}^ka(l_i)\delta_{n,\sum l_i}
\ee
and
\be
B_k(n)=\sum_{m_1\geq 1}\cdots\sum_{m_k\geq 1}\prod_{i=1}^kb(m_i)\delta_{n,\sum m_i}.
\ee
$A_k(n)=B_k(n)=0$ if $n\leq0$ or $k>n$, therefore $Q_{N,n}=0$ for $n\geq N$.
One can show that in the thermodynamic limit
the free energy and the pressure do not depend on the boundary condition.
Below we give these quantities in the case of the simplest model (the LPS model in
a strict sense)
obtained by choosing
\be\label{LPS}
a(l)=a\qquad b(l)=bu^ll^{-\alpha}
\ee
where $u,\alpha>1$ and $a,b>0$ .
With the notations $c=ab$,
\be\label{zetaalpha}
\zeta_\alpha(x)=\sum_{l=1}^\infty l^{-\alpha}e^{-lx} \qquad (x\geq 0)
\ee
and
\be\label{beta-c}
\epsilon_c=\ln u-\ln [1+c\,\zeta_\alpha(0)],
\ee
one obtains
\be
\beta p=\ln u+\Theta(\epsilon-\epsilon_c)x(\epsilon)
\ee
where $\Theta$ is the Heaviside function
and, for $\epsilon>\epsilon_c$,
$x(\epsilon)$ is the unique solution of the equation
\be\label{u}
ue^{x-\epsilon}=1+c\,\zeta_\alpha(x)\ .
\ee
Thus,  $x(\epsilon_c)=0$ and $x(\epsilon)$ is monotone increasing with $\epsilon$.
For $\epsilon<\epsilon_c$, $\beta p$ takes on its minimum value,
$\ln u$. Comparison with (\ref{f}), (\ref{p}) and (\ref{LPS}) shows that the
corresponding density and free energy are $\rho_\epsilon\equiv 0$ and
$f(\rho_\epsilon)\equiv -\ln u$.
Legendre transformation yields $f(\rho)$ for all $\rho$.
With $\rho_c\equiv\rho_{\epsilon_c}$,
\be\label{free-en}
\beta f(\rho)=
\left\{\begin{array}{ll}
\!-\ln u+\epsilon_c\rho,&\rho\leq\rho_c\\
\!-\ln u-y+\rho\!\left(y+\ln\frac{u}{1+c\,\zeta_\alpha(y)}\right),
& \rho\geq \rho_c
\end{array}\right.
\ee
where $y=y(\rho)$ is the solution of the equation
\be\label{ro}
1+\frac{c\,\zeta_{\alpha-1}(y)}{1+c\,\zeta_\alpha(y)}=\rho^{-1}.
\ee
In fact, $y(\rho)=x(\epsilon_\rho)$; $y(\rho)$
is defined for $\rho_c\leq\rho\leq 1$ and
is monotone increasing, $y(\rho_c)=0$ and $y(1)=\infty$. Therefore,
$f(\rho)$ is monotone increasing from $f(0)=-\beta^{-1}\ln u$ to $f(1)=0$;
as the Legendre transform
of the convex $p(\epsilon)$, $f(\rho)$ is also convex. Substituting $y=0$ into
(\ref{ro}) yields
\be\label{roc}
\rho_c=\frac{1+c\,\zeta_\alpha(0)}{1+c\,[\zeta_{\alpha-1}(0)+\zeta_\alpha(0)]}.
\ee
This shows
that $\rho_c=0$ if $\alpha\leq 2$ and $\rho_c>0$ if $\alpha>2$. The first case
corresponds to a continuous phase transition with the order parameter $\rho_\epsilon$
vanishing continuously as $\epsilon$ approaches $\epsilon_c$ from above.
In the second case the transition
is of first order: $\rho_\epsilon$ decreases continuously to $\rho_c$ as
$\epsilon$ decreases to $\epsilon_c$ and jumps to zero below $\epsilon_c$.
Computation of $\rho'_\epsilon=\d\rho_\epsilon/\d\epsilon$
from Eq.~(\ref{ro}) shows that
$\rho'_{\epsilon_c+0}=\infty$ if $\alpha\leq3$ (while it is finite for $\alpha>3$).
The type of divergence can be extracted from the asymptotic forms
\bea\label{rhoasymp}
\rho_\epsilon\sim (\epsilon-\epsilon_c)^{2-\alpha}\quad {\rm if\ }1<\alpha<2
\nonumber\\
\rho_\epsilon-\rho_c\sim\left\{\begin{array}{cl}
(\epsilon-\epsilon_c)^{\alpha-2}&{\rm if\ } 2<\alpha<3\\
|\ln(\epsilon-\epsilon_c)|^{-1}&{\rm if\ }\alpha=2
\end{array}\right.
\eea
valid when $\epsilon$ decreases to $\epsilon_c$. Equation (\ref{rhoasymp}) is
obtained from the asymptotic form of $\zeta_\alpha(x)$ for small $x$ and from
$x(\epsilon)\approx\epsilon-\epsilon_c$ for $\epsilon-\epsilon_c\ll 1$. It will
be used in Section \ref{OPE}. Finally, we note that for $\epsilon<\epsilon_c$ ($T>T_c$),
when $\rho_\epsilon=0$, the decay of the density with increasing $N$ can be
computed,
\be
\rho_N\sim\frac{2}{(1-e^{-\epsilon_c(1-T_c/T)})N}\ .
\ee
This formula is valid for $N\gg(1-T_c/T)^{-1}$.

\section{Correlation functions}

We will compute the one- and two-point correlation functions of
H-bonds in the grand-canonical ensemble for the LPS model.
Let us denote the partition function of the box $[x,y]$ by
\be\label{defin1}
Q^{\alpha \,\beta} [x,y] = Q^{\alpha \,\beta} (y-x+1)
\ee
where we have imposed the boundary conditions $\alpha$ at $x$
and $\beta$ at $y$ (Note that $ Q^{\alpha \,\beta} = Q^{\beta \, \alpha} $,
$ Q^{0 \, 0} (1) =  b u$ and $ Q^{1 \, 1} (1) =  a e^{\epsilon } $).
Without prejudice of generality we
compute the correlation functions with the boundary conditions
$\alpha = 1 $ and $\beta = 1$. We begin with the density
$\rho_N(r)$. In a configuration the site $r$ is in a cluster of 1's
beginning at a site $x+1$, ending at a site $y-1$, so that at sites
$x$ and $y$ we have 0's. It follows that
\be\label{rho1}
Q^{1 \, 1} (N) \rho_N(r) =
a \sum_{y=r+1}^{N-1} \sum_{x=2}^{r-1} Q^{1 \,0} [1,x]
e^{\epsilon (y-x-1)}
Q^{0 \,1} [y,N]
\ee
By introducing
\be\label{defin2}
g(N) = \sum_{x=2}^N  Q^{0 \,1} (x) e^{-\epsilon x}
\ee
we can rewrite (\ref{rho1}) as follows:
\be\label{rho1n}
\rho_N(r) = \frac{ae^{\epsilon N} [g(r-1) + 1][g(N-r) +1]}{Q^{1 \, 1} (N)}.
\ee
For the two-point correlation function $\rho_N(r,r')$ we observe that
in a configuration either $r$ and $r'$ are in the same cluster of 1's or
they belong to  distinct clusters of 1's. By assuming $r' \ge r+1$ we obtain
\bea\label{rho2}
\rho_N(r,r') = \frac{a e^{\epsilon N}[g(r-1) + 1][g(N-r') +1]}
{Q^{1 \, 1} (N)} \times \nonumber\\
\left[1 + a\sum_{l = 1}^{r' -r } Q^{0 \, 0} (l)
e^{-\epsilon l} (r' - r -l) \right]
\eea
These expressions can be simplified by considering that the following
relations hold for $N \ge 2$:
\bea\label{recur}
Q^{0 \, 1} (N) &=& u^{N+1} e^{-\epsilon} Z_N - u^N Z_{N-1}
\nonumber\\
Q^{0 \, 0} (N) &=& \frac{u^N}{a} \left[u^2  e^{-2 \epsilon} Z_{N+1}
- 2u e^{-\epsilon} Z_N + Z_{N-1}
\right]
\nonumber\\
Q^{1 \, 1} (N) &=& a u^N  Z_{N-1}
\eea
where
\be\label{zetan}
Z_N = \frac{1}{2\pi i} \oint_C \frac{{\rm d} w}{w^{N+1}} \,\, \frac{1}{u e^{-\epsilon} - R(w)} \quad ,
\ee
$C$ is some circle around the origin and
\be\label{rofw}
R(w) = w + c \sum_{l=1}^{\infty} l^{-\alpha}
w^{l+1}
\ee
Upon these results we have
\be\label{gnp1}
g(N)+1 = u^{N+1} e^{-\epsilon (N+1)} Z_N
\ee
and finally
\bea\label{corr1}
\rho_N(r) &=& u  e^{-\epsilon} \frac{Z_{r-1} Z_{N-r}}{Z_{N-1}} \\
\rho_N(r,r') &=&  u^2 e^{-2\epsilon} \frac{Z_{r-1} Z_{N-r'} Z_{r'-r}}{Z_{N-1}}
\eea
where $r' \ge r$.

In the low--temperature phase ($\epsilon > \epsilon_c$) we can perform the
thermodynamic limit ($N \to \infty$) by fixing $|r-r'|$ and letting the distance of $r$
to the boundaries also tend to infinity. Making use of the asymptotic ($N\gg1$) relation
\be\label{zetaq}
Z_N \sim \lambda^N q(1 + e^{-\kappa N})
\ee
with $\kappa > 0$,
\be\label{lambdam1}
\lambda^{-1} = \frac{e^{\epsilon}}{u} \, w^{*}
\ee
and
\be\label{qrel}
q^{-1} = w^{*}  \frac{{\rm d} R(w)}{{\rm d} w}|_{ w = w^{*}}\ ,
\ee
where $w^{*} $ is the positive solution of the equation $ u e^{-\epsilon} = R(w^{*})$,
we obtain
\be\label{corr2}
\rho(r) = \rho=u e^{-\epsilon} q = \partial \beta p/\partial \epsilon,
\ee
the expected density of H-bonds, and
\be
\rho(r,r') =  q\, u^2\, e^{-2\epsilon} \lambda^{-|r'-r|} Z_{|r'-r|}.
\ee
We can see that the truncated correlation function $\rho^T(r,r') = \rho(r,r')
- \rho(r) \rho(r')$ decays exponentially
\be\label{rhoT}
\rho^T(r,r') \sim \rho^2 \, e^{-\kappa |r'-r|}
\ee
for $|r' - r | \to \infty$. We note that translation invariance is lost in the
high--temperature phase,
because there remains an explicit dependence on $r$ and $r'$ in $\rho(r,r')$,
even if $r$ and $r'$ are far from the boundaries.
This can be considered as a weakness of the LPS model.
Nonetheless, we shall see that we can still attribute a clear meaning to the
structure factor in this phase.

\section{Structure factors}\label{Struct}

The structure factor is more accessible to experiments than the corresponding
correlation function. It is defined as follows
\be\label{strfN}
S_N(k) = \frac{1}{N} \sum_{r,r'} \rho_N^T(r,r') e^{i k (r'-r)}.
\ee
We aim at computing the limit $S(k)$ of this quantity as $N \to \infty$.
The two cases $T < T_c$ and $T > T_c$ are
treated separately.

{\sl Below $T_c$}| The correlation functions become translational invariant
for $N \to \infty$, i.e. $\rho_N^T(r,r') \to \rho^T(0, r' - r)$. Moreover
$\rho^T(0,r)$ decays exponentially and we obtain
\be\label{strf1}
S(k) =  \sum_{r} \rho^T(0,r) e^{i k r}.
\ee
Making use of (\ref{zetan}), (\ref{rofw}) and (\ref{rhoT}) we have
\be\label{strf2}
\frac{S(k)}{\rho} =
2{\rm Re} \left[\frac{e^{\delta}}{1 + c \zeta_{\alpha}(\delta)
- e^{i k}[1 + c \zeta_{\alpha}(\delta - i k)]}\right]-1.
\ee
In this expression we have reparametrized $w^{*}$ as
\be\label{repar}
w^{*} = e^{-\delta }.
\ee
Accordingly, $T=T_c$ corresponds to $\delta = 0$.
The expression (\ref{strf2}) depends on the details of the LPS model.
However, we may expect that in the critical region, characterized
by $ |k| << 1$ and $|\frac{T- T_c}{T_c}| << 1$ universal features
emerge and the results depend only on the parameter $\alpha$~. If this
is the case, we can be confident of the predictions of the model.
Let us focus on the analysis of the critical region. The parameter
$\delta$ allows us to define a correlation length $\xi$ through the relation
\be\label{xi}
\xi = \delta^{- 1}.
\ee
Near $T_c$, we find
\be\label{corlen}
\xi =\left\{\begin{array}{ll}
 \xi_0(\alpha) (T_c/T-1)^{-\frac{1}{\alpha -1}} & {\rm if\ }1<\alpha<2\\
 \xi_1(\alpha) (T_c/T-1)^{-1} &{\rm if\ } 2<\alpha<3.
 \end{array}\right.
\ee
In order to obtain explicit expressions we need to study the function
$\zeta_{\alpha}(\delta)$, cf. Eq.~(\ref{zetaalpha}),
for small values of $\delta$. For this purpose
we can use the integral representation
\be\label{zeta2}
\zeta_{\alpha}(\delta)
= \frac{1}{\Gamma(\alpha)} \int_0^{+\infty}
\frac{t^{\alpha - 1}}{e^{t+ \delta} -1}\, {\rm d}t
\ee
valid for ${\rm Re\,}\delta >0$. By this relation we can write
\bea\label{zetad}
\lefteqn{
\zeta_{\alpha}(\delta) - \zeta_{\alpha}(\delta -i k)
}\\
&&  = (e^{- i k} -1)
\left[ \zeta_{\alpha -1}(\delta) + (1 - e^{- i k}) B_{\alpha}(k,\delta) \right]
\nonumber
\eea
where
\be\label{b1}
B_{\alpha}(k,\delta) = \frac{1}{\Gamma(\alpha)}\int_0^{+\infty}
\frac{t^{\alpha - 1}}
{(e^{t+ \delta -ik} -1)(1 -  e^{-t -\delta})^2  }\, {\rm d}t .
\ee
For $1 < \alpha < 3$ and $\delta \to 0$ the above expression reduces to
\be\label{b1app}
B_{\alpha}(k,\delta) \approx \delta^{\alpha -3} b_{\alpha}(k)
\ee
with
\be\label{b2}
b_{\alpha}(k) = \frac{\Gamma (2-\alpha)}{\alpha -1}
\left[\frac{(1-ik)^{\alpha -1} -1}{k^2}\right] + i \frac{\Gamma(2- \alpha)}{k}.
\ee
In particular,
\be\label{b0}
b_{\alpha}(0) = \frac{\Gamma (3-\alpha)}{2}.
\ee
The compressibility $\chi = S(0)$ can now be computed. For $1 < \alpha < 2$
the phase transition is continuous and we have
\bea\label{transc}
\rho &\approx& \rho_0 \xi^{\alpha -2}
\\
\chi &\approx& \rho_0^2 (2-\alpha) \xi^{2\alpha - 3}.
\eea
For $2 < \alpha < 3$ the phase transition is of first order and, as
$\rho \to \rho_c$, one has
\be\label{chi}
\chi \approx \chi_0 \xi^{3 - \alpha}.
\ee
We can also find the scaling limit for $S(k)$. It is defined by taking
the limits $k \to 0$ and $\xi \to \infty$ with the product $k \xi$
fixed at some finite value. If $1 < \alpha < 2$ we find
\be\label{scalim}
\frac{S(k)}{S(0)} = \frac{2(\alpha -1)}{2 - \alpha}
{\rm Re} \left[ \frac{1}{(1 - ik \xi)^{\alpha -1} -1} \right].
\ee
For small $k \xi$ this gives a Lorentzian behavior
\be\label{loren}
\frac{S(k)}{S(0)} = \frac{4}{(k\xi)^2 + 4}
\ee
while in the critical region one has
\be\label{crit}
\frac{S(k)}{S(0)} = \frac{2(\alpha -1)\cos \frac{\pi}{2}(\alpha -1)}
{(2 - \alpha)(k\xi)^{\alpha -1}}.
\ee
On the other hand, for $2 < \alpha < 3$ we have
\be\label{strfirst}
\frac{S(k)}{S(0)} = \frac{2\,{\rm Re}\left[ 1 - (1 - ik \xi)^{\alpha -1} \right]}
{(\alpha -1)(\alpha -2)(k\xi)^2}.
\ee
For $|k \xi| <<1$ one has again a Lorentzian, but in the critical region
(\ref{strfirst}) takes the form
\be
\frac{S(k)}{S(0)} =
\frac{2 \, \cos \frac{\pi}{2}(\alpha +1)}{(\alpha -1)(\alpha -2)(k\xi)^{3 - \alpha}}.
\ee

{\sl Above $T_c$} | In this phase the properties of the LPS
model are more dependent on $N$ than in the low-temperature one.
Basically, one has to provide a suitable description of the
partition function $Z_N$, defined in (\ref{zetan}). Let us
introduce the parameter $v = u e^{-\epsilon}$. It can easily be
shown that if $v > v_c$ then
\be\label{zetan2}
Z_N = \frac{1}{2\pi} \int_{-\pi}^{+\pi} e^{-i N \theta} g(\theta)\, {\rm d}\theta
\ee
where
\be\label{gteta}
g^{-1}(\theta) = v - e^{i  \theta} [1 + c \zeta_{\alpha}(-i \theta)].
\ee
In this case the structure factor can be expressed in the form
\be\label{strfac}
N S_N(k) = 2 v^2 {\rm Re\,} C_N(k) - v^2 |D_N(k)|^2 -v D_N(0)
\ee
where
\be\label{cn}
C_N(k) = Z^{-1}_{N-1} \int_{-\pi}^{+\pi}
\frac{{\rm d}\theta}{2\pi}
e^{-i (N-1) \theta}
g^2(\theta)g(\theta + k)
\ee
and
\be\label{dn}
D_N(k) = Z^{-1}_{N-1} \int_{-\pi}^{+\pi}
\frac{{\rm d}\theta}{2\pi}
e^{-i (N-1) \theta}
g(\theta)g(\theta + k).
\ee
In order to find the asymptotic behavior ($N \to \infty$) of
these expressions we use the following formulae. Let $F(x)$
be a function whose only singularity in the interval $\left[-a,a \right]$
is at the origin and the singularity is integrable.
If, close to the origin,
\be\label{Fx}
F(x) \sim \sum_{j} p_j(s) |x|^{r_j} ,\quad s= \sgn(x)
\ee
then
\be\label{IFx}
\int_{-a}^a F(x) e^{-i N x}\d x \sim \sum_{sj} p_j(s) \Gamma(1 + r_j)
(N e^{\frac{i \pi}{2} s}  )^{-(1+ r_j)} \,\, ,
\ee
where $r_j$ is a sequence of real numbers such that
$-1 < r_1 < r_2 < \cdots$ .
Using Eqs.~(\ref{zeta2})--(\ref{b2}),
this formula yields
\be\label{ZeN}
Z_N \sim c/ (\bar\delta^2 N^{\alpha})
\ee
where $\bar\delta = v - v_c$. Moreover, if $k \not= 0$ then
\be\label{cna}
C_N(k) \sim g(k) + g^2(-k) e^{i k (N-1) }
\ee
and
\be\label{dna}
D_N(k) \sim g(k) + g(-k) e^{i k (N-1) }.
\ee
This shows that the compressibility
\be\label{chi2}
\chi=S(0) \sim 2 v v_c/(N \bar\delta^3).
\ee
Accordingly, in the limit $N \to \infty$ we have
\be\label{strfa}
\frac{S(k)}{S(0)} = \frac{v}{v_c}|g(k)|^2 \left[ (2/\bar\delta)
{\rm Re\,} g^{-1}(k) -1\right] - \frac{\bar\delta}{v_c}.
\ee
Since $\bar\delta = \bar\delta_0 (T/T_c -1)$, also in this case
we can obtain the scaling limit. With the definition
(\ref{corlen}) of the correlation length $\xi$,
for $1<\alpha<2$ the normalized structure factor reads
\be\label{strab1}
\frac{S(k)}{S(0)} =
\frac{
1 + 2(k\xi)^{\alpha -1} \sin\frac{\pi}{2}\alpha
}
{
1 + (k\xi)^{2(\alpha -1)} + 2(k\xi)^{\alpha -1} \sin\frac{\pi}{2}\alpha
}
\ee
while, for $2<\alpha<3$,
\be\label{strab2}
\frac{S(k)}{S(0)} =
\frac{1}{1 + (k\xi)^2}.
\ee
We stress that the main result of these computations is the
existence of a universal scaling limit for the normalized structure factor
$S(k)/S(0)$, which depends only on the parameter $\alpha$.
This suggests also that the LPS model belongs to a new universality
class, in the language of critical phenomena. Accordingly, its predictions
about the process of DNA denaturation can be taken with some confidence.
Moreover, we expect that this result may help to clarify experimentally
the nature of the phase transition, either continuous or first--order.

\section{Microscopic dynamics}\label{MiDy}

We will describe the evolution of the configuration $\n$ by a discrete-time local
dynamics. Although the method is well-known, we present it in some detail.

In a time step, $\n$ can change only at a single site $x$. The new
configuration is $T_x\n$ and, hence, $(T_x\n)_y=n_y$ if $y\neq x$, and
$(T_x\n)_x=1-n_x$. Note that $T_x^2$ equals the identity.
The transition probability from $\n$ to $\m$ is $p_{\n,\m}$.
More precisely, it is the conditional probability that,  if the configuration is $\n$
in time $t$, it will be $\m$ in time $t+1$. It then follows that $p_{\n,\m}=0$
unless $\m=\n$ or $\m=T_x\n$ for some $x$. The irreducibility of the matrix
$p_{\n,\m}$ and, hence, the ergodicity of the dynamics and the
uniqueness of the equilibrium state is guaranteed if
$p_{\n,T_x\n}$ is indeed positive for {\em all} $x$. (Note,
however, that both $\n$ and $T_x\n$ have to satisfy the boundary conditions. If
these are given by fixing $n_1$ and $n_N$ then $2\leq x\leq N-1$.)
In a numerical experiment the transition probability is decomposed as
\be
p_{\n,T_x\n}=p_x W(T_x\n|\n,x)
\ee
where $p_x$ is the probability to choose $x$ for an attempt of change, and
$W(T_x\n|\n,x)$ is the conditional probability that once $\n$ and $x$ have been
selected,
the change is executed.
Clearly, $\sum_xp_x=1$ must hold; for periodic or free boundary
conditions the typical choice
for $p_x$ is $1/N$. The probability to stay in $\n$ if the change is attempted
in $x$ is
\be
W(\n|\n,x)=1-W(T_x\n|\n,x)
\ee
and, thus,
\be
p_{\n,\n}=1-\sum_{x=1}^Np_{\n,T_x\n}=\sum_{x=1}^Np_xW(\n|\n,x)\ .
\ee
This is nonzero unless $W(\n|\n,x)=0$ for all $x$.

Let $p_t(\n)$, $t=0,1,2,\ldots$, be the probability of $\n$ at time $t$. The master
equation of the evolution is
\be\label{master}
p_{t+1}(\n)=p_t(\n)p_{\n,\n}+\sum_{x=1}^Np_t(T_x\n)p_{T_x\n,\n}\ .
\ee
The local transition probability $W$ is to be chosen so that the grand-canonical
distribution $\pi$ be the unique equilibrium solution of (\ref{master}). Imposing
the condition of detailed balance
\be
\pi(\n)W(T_x\n|\n,x)=\pi(T_x\n)W(\n|T_x\n,x)\ ,
\ee
we still have an infinity of choices: any $W$ satisfying
\be\label{trans}
\frac{W(T_x\n|\n,x)}{W(\n|T_x\n,x)}=e^{-U(T_x\n)+U(\n)}=z(\n,x)
\ee
and $0<W(T_x\n|\n,x)\leq 1$ for all $x$ and all $\n$ of length $N$
with the right boundary conditions will do.

If $\gamma$ is an interval (a sequence of consecutive integers between 1 and $N$)
then by definition the distance of $x$ to $\gamma$ is
$d(x,\gamma)=\min\{|x-y|:y\in\gamma\}$. Because $U(\n)$ is additive in the
contributions of the intervals of $\n$,
\be
U(\n)=-\sum_i[\ln a(l_i)+l_i\epsilon]-\sum_j \ln b(m_j)\ ,
\ee
$z(\n,x)$ depends only on the intervals of $\n$ and $T_x\n$ whose
distance to $x$ is 0 or 1. If in $\n$ there is a single interval
with this property, there will
be three in $T_x\n$, and {\em vice versa}\,; if there are two in $\n$,
there will also be two in $T_x\n$. We obtain altogether three
different functions of the lengths of
intervals, and their reciprocals. These are listed in Table 1.

\begin{table}[tbh]
\centering
\begin{tabular}{||c|c||} \hline
$(1)^{l_1} 1 (1)^{l_2} \,\to\, (1)^{l_1} 0 (1)^{l_2} $ & $ R_1 $ \\ \hline
$(0)^{l_1} 1 (1)^{l_2} \,\to\, (0)^{l_1} 0 (1)^{l_2} $ & $ R_2 $ \\ \hline
$(1)^{l_1} 1 (0)^{l_2} \,\to\, (1)^{l_1} 0 (0)^{l_2} $ & $ R_3 $ \\ \hline
$(0)^{l_1} 1 (0)^{l_2} \,\to\, (0)^{l_1} 0 (0)^{l_2} $ & $ R_4 $ \\ \hline
$(1)^{l_1} 0 (1)^{l_2} \,\to\, (1)^{l_1} 1 (1)^{l_2} $ & $ R_1^{-1}$ \\ \hline
$(0)^{l_1} 0 (1)^{l_2} \,\to\, (0)^{l_1} 1 (1)^{l_2} $ & $ R_2^{-1}$ \\ \hline
$(1)^{l_1} 0 (0)^{l_2} \,\to\, (1)^{l_1} 1 (1)^{l_2} $ & $ R_3^{-1}$ \\ \hline
$(0)^{l_1} 0 (0)^{l_2} \,\to\, (0)^{l_1} 1 (0)^{l_2} $ & $ R_4^{-1}$ \\ \hline
\end{tabular}
\caption{ Transition rates $z({\bf n},x)$.
The notations $(X)^ {l_1}$ and $(X)^ {l_2}$
indicate a left-cluster of length $l_1$ and a right-cluster
of length $l_2$ made of $X$-symbols, respectively.
$R_1 = ab \, v$,
$R_2 = v\, (\frac{l_1}{l_1 + 1})^{\alpha}$,
$R_3 = v\, (\frac{l_2}{l_2 + 1})^{\alpha}$,
$R_4 = \frac{v}{ab} \, (\frac{l_1 l_2}{l_1+l_2+1})^{\alpha}$
where $v=ue^{-\epsilon}$.}
\label{ratetable}
\end{table}

Typical choices of $W$ are
\be
W(T_x\n|\n,x)=\min\{1,z(\n,x)\}\,\,\,{\rm or}\,\,\, \frac{z(\n,x)}
{1+z(\n,x)}\ .
\ee
They satisfy (\ref{trans}) because $z(T_x\n,x)=1/z(\n,x)$. We shall work with the
first one.

\section{The order parameter equation}\label{OPE}

The microscopic dynamics of denaturation as described by the master equation is
quite complex and untractable analytically. Most of the time, however, one is
interested in the evolution of the order parameter, namely the density $\rho$ of
1 (H-bonds). This quantity, obtained as an average over the sample, should vary
much more slowly than the other degrees of freedom to which it is coupled. This
coupling has two effects. (i) It creates an effective thermodynamic force
exerted on the order parameter and depending nonlinearly on it. This force should
vanish if the density, and via it the pressure, reach their equilibrium value,
corresponding to the prescribed value of $\epsilon$.
We choose it therefore in the form
$$-\left[(\beta f)'(\rho)-\beta\mu\right]$$
where $f(\rho)$ is the {\em equilibrium} free energy for each value of $\rho$,
cf. Eqs.~(\ref{free-en}). So the deterministic part of the dynamics realizes the
search of the supremum in Eq.~(\ref{L1}).
One can recognize here also the mean-field approach to the problem of
the critical slowing down in phase transitions. (ii)
Due to the very
large number of the coupled degrees of freedom,
the coupling
creates a random force of
zero average, responsible for the fluctuations of the density.
We choose the random force as a
white noise $\eta(t)$ with a correlation
\be
\langle\eta(t)\eta(t')\rangle=2\gamma\beta^{-1}\delta(t-t').
\ee
Thus, the time evolution of $\rho(t)$ is
governed by the Langevin equation
\be\label{rhodyn}
\dot{\rho}=-\gamma[(\beta f)'(\rho)-\beta\mu]+\sqrt{\beta}\ \eta(t)
\ee
where $\gamma>0$.
We expect this equation to describe correctly the behavior of the order
parameter in the vicinity of the transition point. The comparison of the analytical
predictions based on Eq.~(\ref{rhodyn}) with the numerical solution of the master
equation can be a first check of this hypothesis. From now on we consider only
the LPS case (\ref{LPS}). From Eq.~(\ref{free-en}) we deduce
\be
(\beta f)'(\rho)=\epsilon_\rho,
\ee
therefore (\ref{rhodyn}) reads
\be\label{dyn}
\dot{\rho}(t)=-\gamma[\epsilon_{\rho(t)}-\epsilon]+\sqrt{\beta}\ \eta(t)
\ee
As we have seen in the former section, the microscopic dynamics depends only on
$\alpha$, $c=ab$ and $v$. 
It is important to realize that $\epsilon_\rho-\epsilon$ also depends only on these
parameters.
Indeed, for given $\rho$ we solve Eq.~(\ref{ro}) for $y$, replace $x$ by $y(\rho)$
in Eq.~(\ref{u}) and extract $\epsilon_\rho$. Thus,
\be
\epsilon_\rho-\epsilon=y(\rho)-\ln \{1+c\zeta_\alpha[y(\rho)]\}+\ln v.
\ee
On the other hand, we expect to obtain the value of $\gamma$ from the microscopic
dynamics.

Note that $\epsilon_\rho\equiv\epsilon_c$ if $\rho\leq\rho_c$ and tends to $\infty$ as
$\rho$ approaches 1. Inverting $\rho_\epsilon$ shows
that $\epsilon_\rho$ starts with zero derivative if $\alpha\leq 3$,
while $\epsilon'_{\rho_c+0}>0$ if $\alpha>3$. We will confine ourselves to the case
$\alpha\leq 3$.

\vspace{2mm}
\noindent
\subsection{Evolution without noise}

\noindent
{\em CASE 1:}$\quad 2<\alpha<3$.
In this case $\rho_c>0$. 
The evolution is naturally different above and below the critical temperature.

\noindent
(i) $T>T_c$ ($\epsilon<\epsilon_c$). Let $\rho_0=\rho(0)$.
We distinguish between two subcases.

\vspace{1mm}
\noindent
1. $\rho_0<\rho_c$. Then $\dot{\rho}=-\gamma\epsilon_c(1-T_c/T)$ and
\be
\rho(t)
=\rho_0-\gamma\epsilon_c(1-T_c/T)t\quad{\rm for}\quad 0\leq t\leq t_f
\label{rhot}
\ee
where $t_f=\rho_0/\gamma\epsilon_c(1-T_c/T)$ and $\rho(t_f)=0$.

\vspace{1mm}
\noindent
2. $\rho_0>\rho_c$. Then for $0\leq t\leq t_c$, $\rho$ evolves according to
$\dot{\rho}=-\gamma(\epsilon_\rho-\epsilon)$, where $t_c$ is defined by
$\rho(t_c)=\rho_c$. For $t>t_c$ the force is constant as in 1., hence
\be
\rho(t)=\rho_c-\gamma\epsilon_c(1-T_c/T)(t-t_c)\quad{\rm for}
\quad t_c\leq t\leq t_f
\label{rhot_l}
\ee
where $t_f=t_c+\rho_c/\gamma\epsilon_c(1-T_c/T)$ and $\rho(t_f)=0$.

What we see here is
{\em the phenomenon of gelation:} The system reaches equilibrium in a
finite time \cite{ziff}.

\vspace{1mm}
\noindent
(ii) $T<T_c$ ($\epsilon>\epsilon_c$). Then $\lim_{t\to\infty}\rho(t)=
\rho_\epsilon>\rho_c$. For large $t$, by linearizing $\epsilon_\rho$ about
$\epsilon$ we find $\epsilon_\rho-\epsilon\approx(\rho-\rho_\epsilon)/\rho'_\epsilon$
[recall that $ \rho'_\epsilon=\d\rho_\epsilon/\d\epsilon$]
and
\be\label{exp1}
\rho(t)-\rho_\epsilon\sim Ae^{-[\gamma/\rho'_\epsilon]t}.
\ee
When $T$ is close to $T_c$, from Eq.~(\ref{rhoasymp}) we deduce
$\rho'_\epsilon\sim (\alpha-2)\epsilon_c(T_c/T-1)^{\alpha-3}$ and
\be\label{exp2}
\rho(t)-\rho_\epsilon\sim\exp(-t/\tau)
\ee
with
$$
\tau=\rho'_\epsilon/\gamma\sim\gamma^{-1}(\alpha-2)(T_c/T-1)^{\alpha-3}.
$$

\vspace{1mm}
\noindent
{\em CASE 2:}$\quad 1<\alpha<2$, $\rho_c=0$.

\noindent
(i) $T>T_c$. Now starting from $\rho_0=\rho(0)>0$, for all times
$\epsilon_\rho\geq\epsilon_c>\epsilon$, therefore
$\dot{\rho}\leq -\gamma\epsilon_c(1-T_c/T)$ and
$\rho(t)$ attains zero in a finite time
$t_f<\rho_0/\gamma\epsilon_c(1-T_c/T)$.
Again, we find gelation.

\vspace{1mm}
\noindent
(ii) If $T<T_c$, for large times
we find the same result as in (\ref{exp1}) and
(\ref{exp2}) with
\[
\tau=\rho'(\epsilon)/\gamma\sim \gamma^{-1}(2-\alpha)(T_c/T-1)
^{1-\alpha}.
\]

\vspace{1mm}
\noindent
{\em CASE 3:}$\quad \alpha=2$. For $T>T_c$ we obtain gelation,
for $T<T_c$ an exponential relaxation to $\rho_\infty=\rho_\epsilon$
with decay time
$$\tau\sim [\gamma(T_c/T-1)\ln^2(T_c/T-1)]^{-1}.$$

\vspace{1mm}
As in the case of the static structure factor, we can obtain a scaling limit
for the time-dependent order parameter $\rho(t)$. It can be determined
by taking $\rho\to 0$, $t\to\infty$, $T\to T_c$.
We restrict this analysis to the case $\alpha>3/2$.

\noindent
(i) $T > T_c$. The gelation time $t_f$ diverges as
$$t_f \sim (T/T_c -1)^{-(1-\nu)} \,\,
{\rm where }\quad
\nu=\frac{2-\alpha}{\alpha-1}.
$$

\noindent
(ii) $ T < T_c$. Then
\be
\frac{\rho(t)}{\rho_\epsilon}=z\left(
\frac{\gamma\epsilon}{\rho_\epsilon}
(1-T/T_c)\,t \right)
\ee
where $z(t)$ is the unique solution of
\be
\dot{z}=-(z^{1/\nu}-1)
\ee
tending to 1 as $t$ tends to infinity.
Therefore, if
$t(1-T/T_c)^{1-\nu}\gg 1$ then
\be
\rho(t) \sim {\rho_\epsilon}\left[1+\exp\left(-\frac{\gamma\epsilon
}{\nu \rho_\epsilon} (1-T/T_c)\,t \right)\right].
\ee
On the other hand, in the critical region
$$1 \ll t \ll |T/T_c -1|^{-(1-\nu)}$$
one has
\be
\rho(t)\sim \left[\frac{\nu\gamma\epsilon t}{1-\nu}
\right]^{-\nu/(1-\nu)}.
\ee

\subsection{The effect of noise}

In order to compute the time--dependent correlation functions of the
density $\rho$, we have added a white noise term $\eta(t)$ to the
deterministic order parameter equation (\ref{rhodyn}). On the other
hand, this choice may lead to a consistency problem. In fact, it is well
known that in this case the stationary distribution ${\mathcal P}(\rho)$
of the density is of the form
\be\label{statdistr}
{\mathcal P}(\rho) \sim \exp\left\{- \frac{1}{2}[\beta f(\rho) - \epsilon \rho]\right\}.
\ee
The average of $\rho$ taken with ${\mathcal P}(\rho)$ is, in general, different
from the equilibrium value $\rho_\epsilon$,
given by the equation
\be\label{rhostar}
(\beta f)'(\rho_\epsilon)=\epsilon,
\ee
cf. Eq.~(\ref{L1}).
This is a direct consequence of having interpreted $f(\rho)$ in (\ref{rhodyn})
as the equilibrium free energy. Therefore, we will only investigate the effect of
noise by considering the linearized version of Eq.~(\ref{rhodyn})
close to equilibrium, i.e.
\be\label{rdlin}
\dot{\rho}=-\gamma(\beta f)''(\rho_\epsilon)
(\rho - \rho_\epsilon) + \sqrt{\beta}\eta(t).
\ee
Below $T_c$, $f''(\rho)$ is related to the compressibility,
\be\label{sta}
\frac{\partial^2 f}{\partial \rho^2} = k_B T/\chi
\ee
which is a well defined quantity in the limit $N \to \infty$. This is not
the case above $T_c$, where the deterministic equation predicts the
occurrence of gelation. In order to circumvent this difficulty, we assume
that (\ref{rdlin}) and (\ref{sta}) are still valid above $T_c$,
while we maintain the explicit dependence on $N$ of both $\rho_\epsilon$ and
$\chi$. The time--dependent correlation function $C_{\rho} (t)$ is
defined as
\be\label{corfut}
C_{\rho} (t) = \langle \rho(t+t') \rho(t')\rangle - \langle\rho(t+t')\rangle
\langle\rho(t')\rangle
\ee
where the bracket denotes the time average. The power spectrum of $\rho$
is given by
\be\label{powsp}
P_{\rho}(\omega) =
 \frac{1}{2\pi} \int_{-\infty}^{+\infty}  e^{i \omega t} C_{\rho} (t)\,{\rm d} t.
\ee
Making use of (\ref{rdlin}) and (\ref{sta}) one obtains
\be\label{corfut1}
C_{\rho} (t) = \chi\, e^{ - |t|/\tau }
\ee
and
\be\label{powsp1}
P_{\rho}(\omega) = \frac{\chi\tau}{\pi} \frac{1}{1 + (\omega \tau)^2}
\ee
where $\tau = \chi/\gamma $.
For $T \nearrow T_c$ we obtain
\be\label{tau}
\tau\sim\left\{\begin{array}{ll}
  \gamma^{-1}\xi^{2\alpha - 3} & {\rm if\ }\quad 1<\alpha<2\\
  \gamma^{-1} \xi^{3 - \alpha} & {\rm if\ }\quad  2<\alpha<3
 \end{array}\right.
\ee
with $\xi$ defined in Section \ref{Struct}.
In the language of critical phenomena 
Eq.~(\ref{tau}) can be interpreted as a scaling
law valid for $\omega \tau$ not too large. The dynamical critical exponent
$z$ is then given by
\be\label{zetae}
z  = \left\{\begin{array}{ll}
  2\alpha - 3 & {\rm if} \quad 1< \alpha < 2\\
  3 - \alpha  &  {\rm if} \quad 2< \alpha < 3\ .
\end{array}\right.
\ee
Near to but above $T_c$, one has instead
\be\label{tauab}
\tau \sim  \gamma^{-1} \xi^2/N
\ee
since $\chi = 2 v v_c/N {\bar\delta}^3$,  see (\ref{chi2})~.

\subsection{Comparison with the numerical experiment}

We have performed numerical simulations according to the microscopic
MonteCarlo dynamics described in Section \ref{MiDy},  in order to check its
agreement with the order--parameter equation. 
We report here the results obtained for the two cases 
$\alpha = 2.5$ (first--order
phase transition) and $\alpha = 1.5$ (second--order phase transition). 
The microscopic dynamics is found to depend on 
the control parameter $v = u e^{-\epsilon}$. Since $\epsilon = \beta \mu $,
$v$ can be used at the place of the temperature $T \sim \beta^{-1}$ for 
describing numerical results.
Without prejudice of generality one can fix 
$ab=1$. According to Eq. (\ref{beta-c}) the critical value $v_c$ is given 
by the relation
\be
v_c = 1 + \zeta_\alpha(0)
\ee
Most of the results discussed in this section have been obtained for
lattice size $N= 5 \times 10^3$. The time evolution of $\rho(t)$ has been
extended over time spans ranging between $ 100 \tau$ and $600 \tau$.
In this subsection $\tau = N$ denotes the natural time unit of lattice updates
(not to be confused with the $\tau$ used in Eq.(\ref{corfut1})~)
~.
Fluctuations have been smoothened by averaging over a large number of
initial conditions (typically $10^4$). The initial conditions have been
sampled among high density equilibrium states obtained for $v = 0.1$
($\rho(0) \approx 0.9$). For the special situation concerning the case
$\alpha = 2.5$ and $\rho(0) < \rho_c$ the initial conditions have been
sampled by attributing to each site the value "1" with probability
$\rho(0)$ and "0" with probability $(1-\rho(0))$~.
Lacking an equilibrium state of finite density
due to the first--order nature of the phase transition, this is
a natural choice for sampling states of fixed initial density $\rho(0)$.
We have also verified that other recipes can modify the duration of the
transient evolution,
but we have not observed any difference in the long time behavior.
Finally, we have also verified that the results do not depend on the 
choice of boundary conditions. In particular, the results reported
in this subsection have been obtained for fixed boundary conditions,
where the first (last) lattice site is put in contact with
a fixed "0" state on its right (left). 

\vspace{1mm}
\noindent
{\em 1. The case} $\alpha = 2.5$

\begin{figure}[h]
\includegraphics[clip,width=6.5cm]{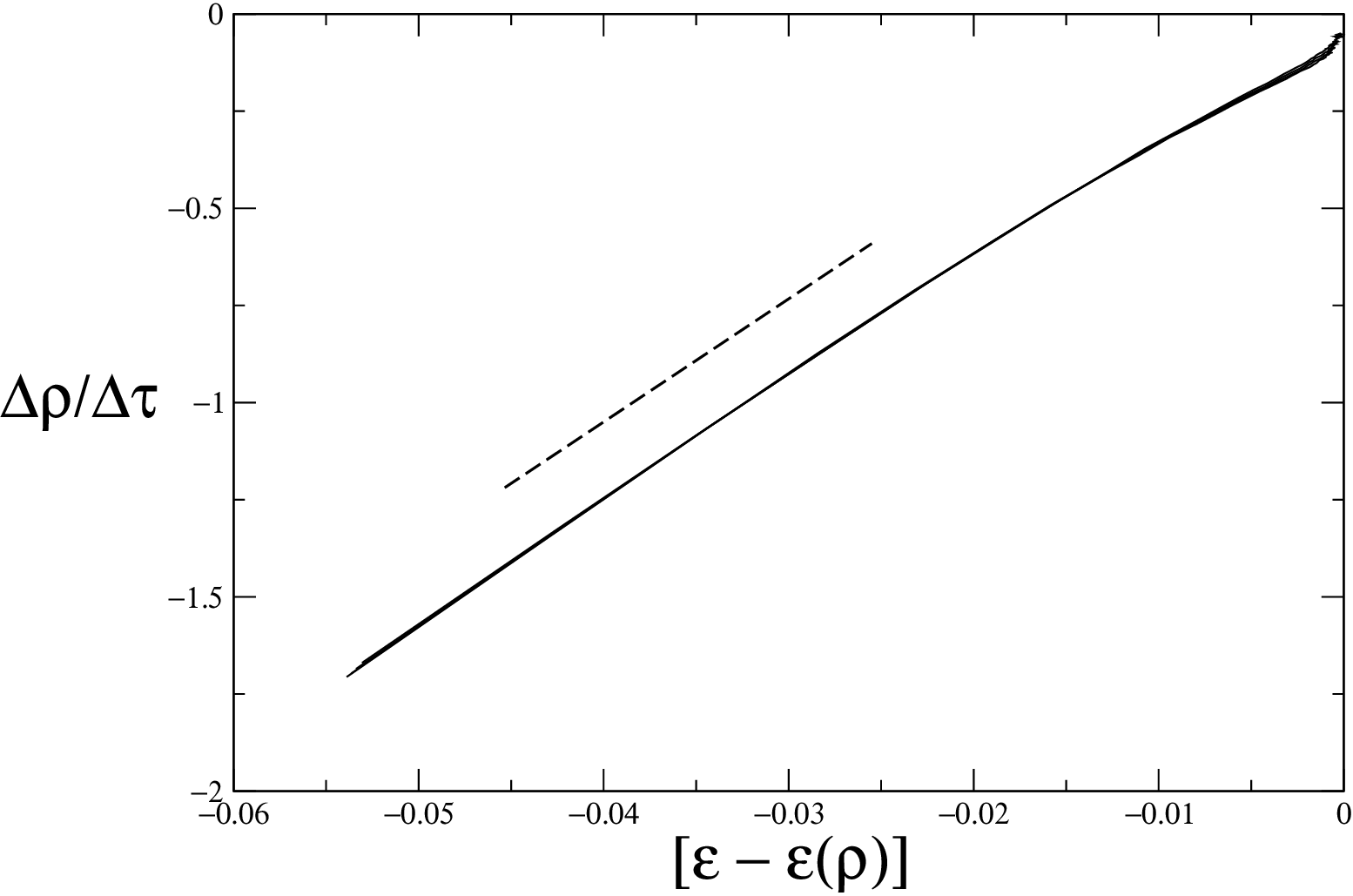}
\caption{
\small
$\dot\rho \equiv \Delta\rho/\Delta\tau$ versus $\epsilon -
\epsilon(\rho)$ for $\alpha =2.5$ and $v = 2.4,\, 2.45,\, 2.5,\, 2.53$.
The dashed line is the best--fit of the slope ($-\gamma
\approx 30$),  common to
all values of $v$ . It has been drawn to guide the eyes for 
appreciating the extension of the transient evolution, during
which the four lines overlap.
}
\label{fig1}
\end{figure}

The time evolution of $\rho(t)$ has been analyzed for several values of
$v$ in the range $[2.0 , 3.0]$, starting from high density equilibrium
states. Close to the theoretical critical value, $v_c = 2.341486...$,
the dynamics has been sampled over a time lapse up to $600\tau$, in order
to obtain a reliable identification of the critical point. Despite finite
size effects are expected to affect significantly MC dynamics in models
with long--range interactions like the LPS model, we have been able to
obtain the numerical estimate $v_c= 2.34 \pm 0.02$, which agrees very well
the theoretical expectation. 
Moreover, for $v > v_c$ (i.e. $T > T_c$) we have also verified that
$\rho (t)$ initially evolves according to the dynamics $\dot\rho =
- \gamma [\epsilon_{\rho} -\epsilon]$ (see Fig.\ref{fig1}). 

\begin{figure}[h]
\includegraphics[clip,width=6.5cm]{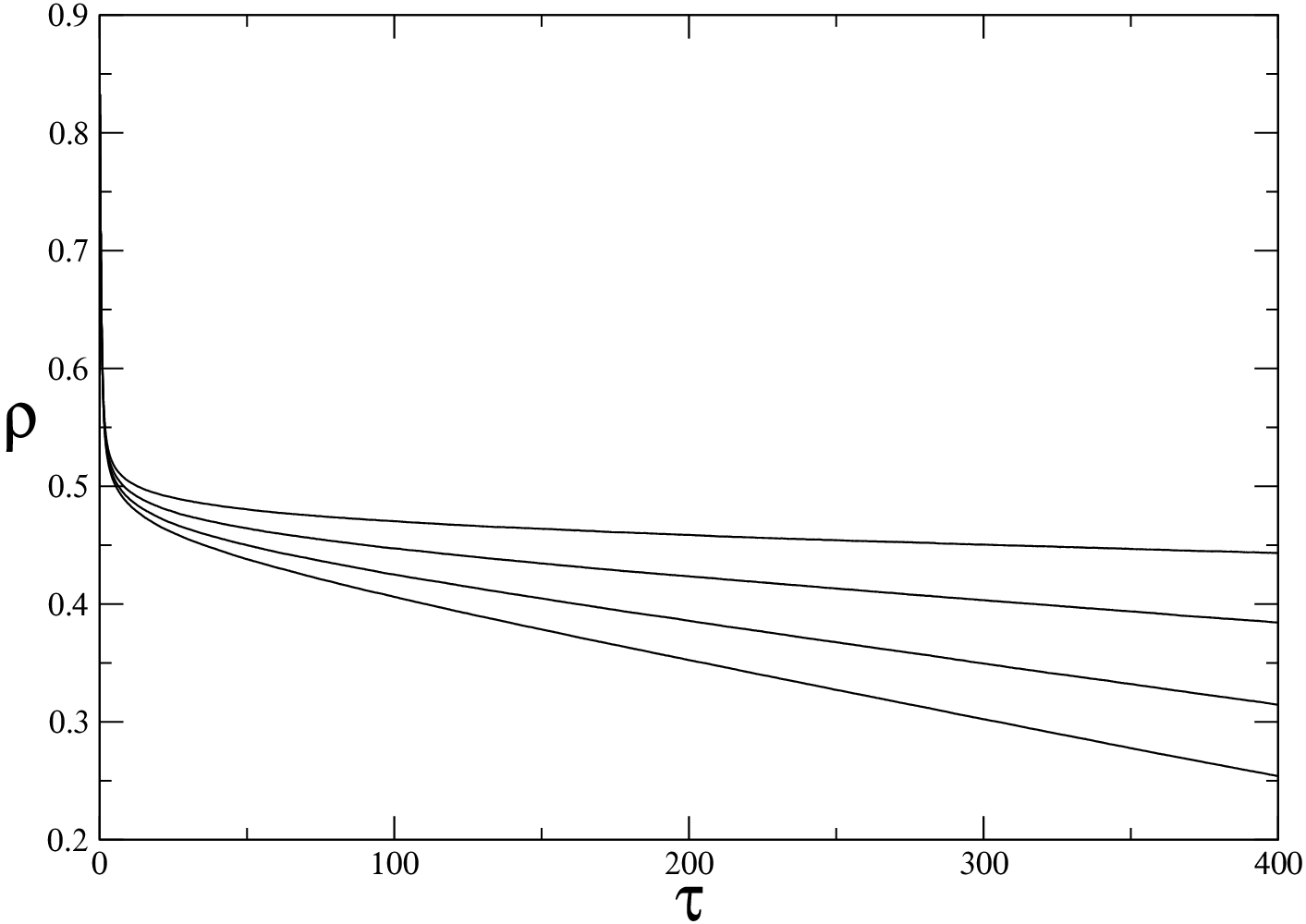}
\caption{
\small
The density of H-bonds $\rho$ as a function of $\tau$ for 
$\alpha =2.5$ and $v
= 2.4,\, 2.45,\, 2.5,\, 2.53$ (from top to bottom). Note that
a linear decrease in time is eventually approached, thus yielding
denaturation after a finite time ({\rm gelation phenomenon}).
}
\label{fig2}
\end{figure}

For larger values of
time, $\rho(t)$ decreases linearly in time, as predicted by 
Eq.(\ref{rhot_l}). This confirms the presence of the phenomenon
of gelation, according to which the system reaches a denutared state in
a finite time (see Fig.\ref{fig2}). 

We have not been able to check the 
quantitative agreement with the theoretical predictions reported in
subsection A. In particular, we have not found an effective criterion for 
locating
the crossover between the initial and the asymptotic dynamics of $\rho(t)$.
It is expected to occur at the critical time $t_c$ ($\rho(t_c)=\rho_c$), but
in MC simulations the crossover region extends over a long time 
lapse, which typically begins before and extends far beyond $t_c$.
We want to point out that such a scenario is not unexpected: finite-size
and memory effects cannot allow for the identification of a single transition 
point between the two dynamical regimes. This analysis could be slightly 
improved by considering much larger lattices and longer simulations, while 
maintaining at least the same number of averages over initial conditions. 
We did not proceed along this line, because the limit of our 
computational capabilities has already been reached with the values adopted 
in the reported simulations.

\begin{figure}[h]
\includegraphics[clip,width=6.5cm]{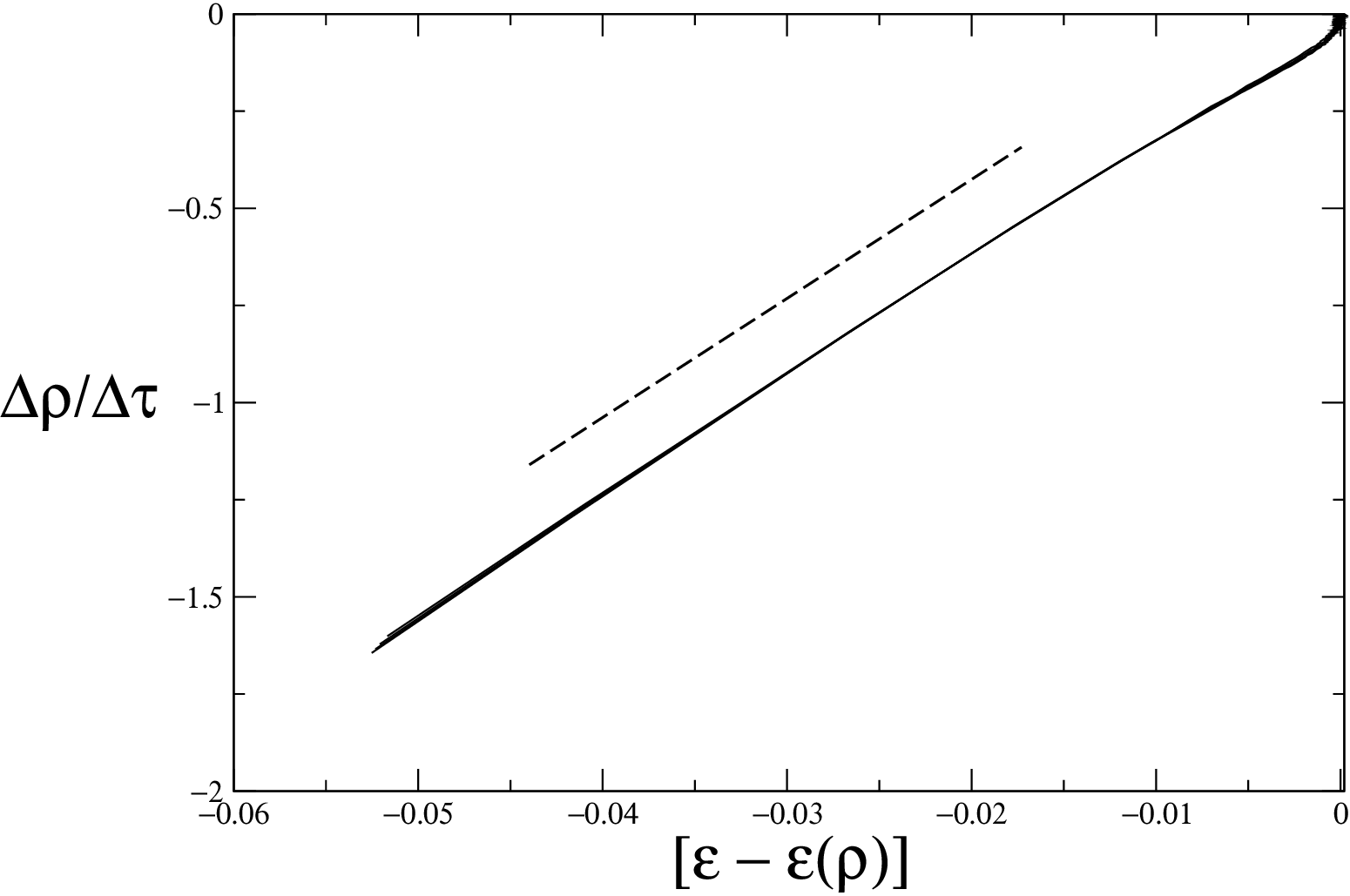}
\caption{
\small
$\dot\rho \equiv \Delta\rho/\Delta\tau$ versus $\epsilon -
\epsilon(\rho)$ for $\alpha =2.5$ and $v = 2.2,\, 2.25,\, 2.3,\, 2.33$.
The dashed line is the best--fit of the slope ($-\gamma
\approx 30$),  common to
all values of $v$ . It has been drawn to guide the eyes for 
appreciating the extension of the transient evolution, during
which the four lines overlap.
}
\label{fig3}
\end{figure}

This scenario does not change also for $v < v_c$ ($T < T_c$)~. 
We still find a 
qualitative agreement with the theoretical predictions based on the
order parameter dynamics. Specifically, for small times the dynamics
is again ruled by $\dot\rho = -\gamma [\epsilon_\rho -\epsilon]$
(see Fig.\ref{fig3})~.

\begin{figure}[h]
\includegraphics[clip,width=6.5cm]{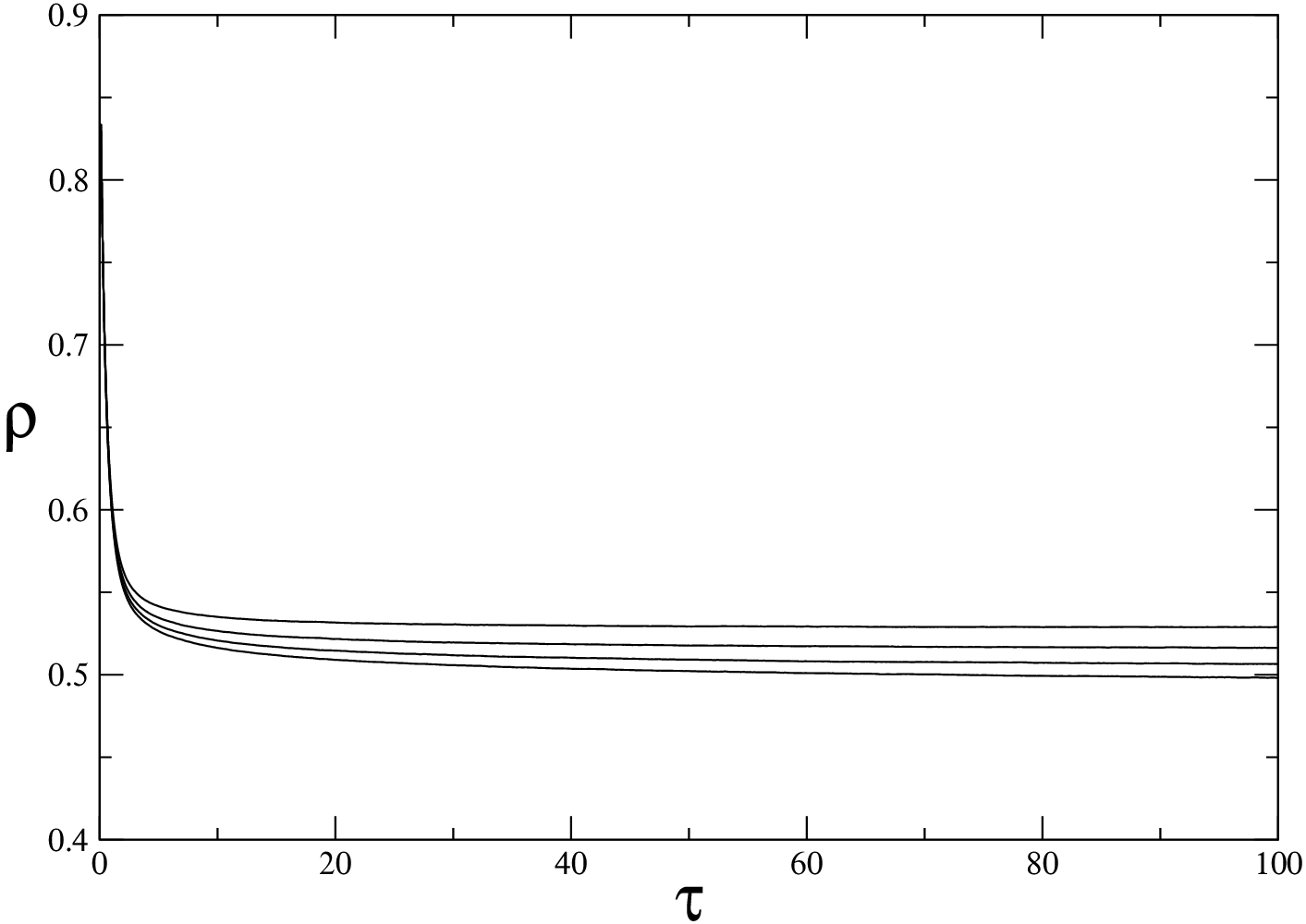}
\caption{
\small
The density of H-bonds $\rho$ as a function of $\tau$ for 
$\alpha =2.5$ and $v
= 2.2,\, 2.25,\, 2.3,\, 2.33$ (from top to bottom). Note that
the decay to a finite value $\rho_{\epsilon}$ slows down significantly
as $v$ approaches $v_c= 2.34 \pm 0.02$.
}
\label{fig4}
\end{figure}

Afterwards, one observes the decay to a finite asymptotic value
$\rho_{\epsilon} > \rho_c$ (see Fig.\ref{fig4})~.
According to Eq.(\ref{exp1}) such a decay is 
predicted to be exponential, with the decay rate $\tau$ given in
Eq.(\ref{exp2})~.

Also in this case a clear quantitative verification is prevented by the
long lapse of time characterizing the crossover between the transient
and the asymptotic dynamics.

\begin{figure}[h]
\includegraphics[clip,width=6.5cm]{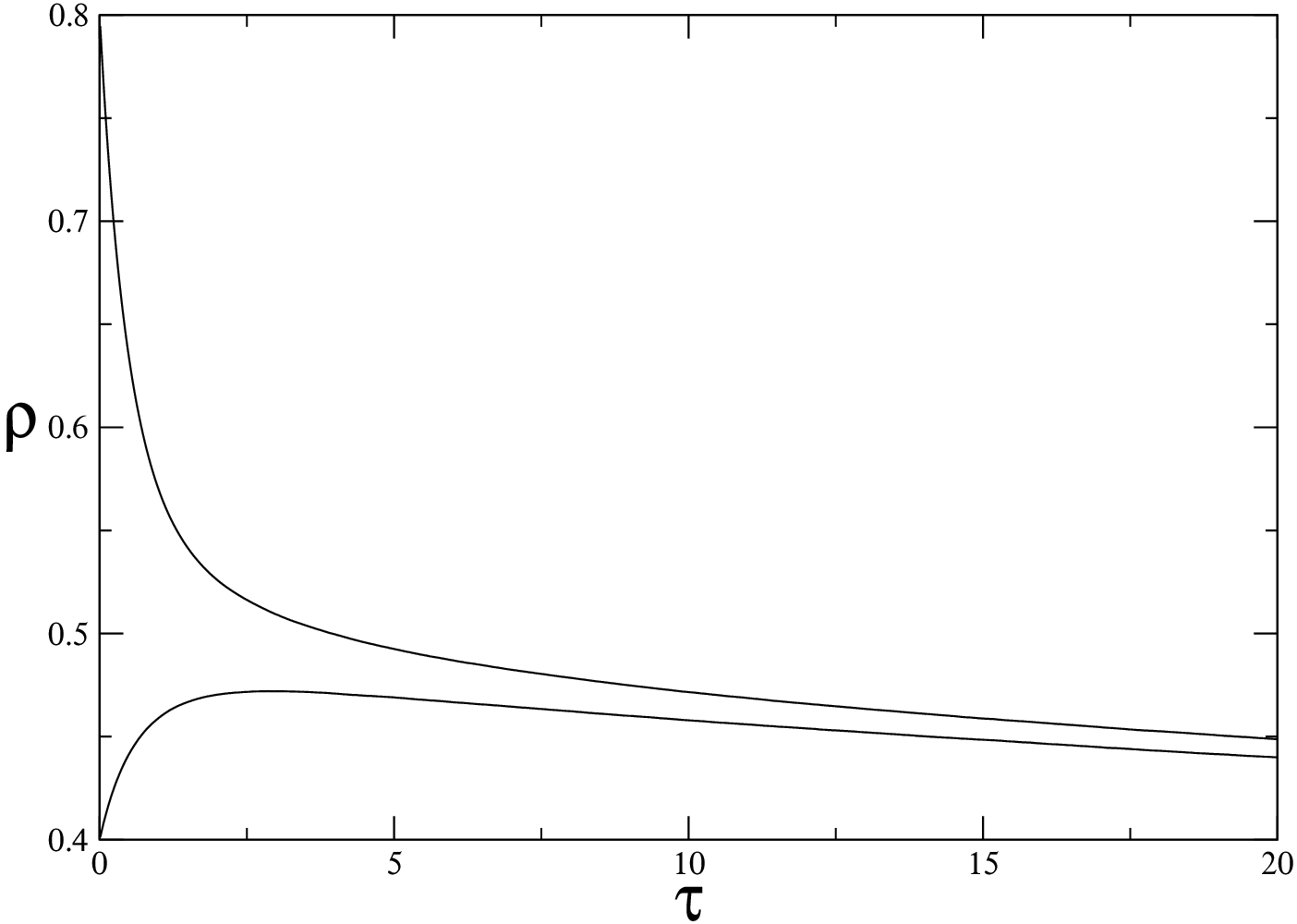}
\caption{
\small
The density of H-bonds $\rho$ as a function of $\tau$ for 
$\alpha =2.5$ and $v = 2.6$. The evolution has been obtained by
averaging over randomly seeded initial conditions of initial
density $\rho(0) = 0.8 $ and 0.4~. In order to better appreciate
the hysteretic effect associated with the first--order nature of 
the phase transition (lower curve), we have reported the evolution 
over a relatively short time scale. Note also that 
the initial conditions affect only the
transient evolution: both curves converge to the same asymptotic
dynamics, which eventually turns out to a linear--in-time decay,
yielding the gelation phenomenon.
}
\label{fig5}
\end{figure}

Further peculiar behaviors of the microscopic dynamics emerge for
$v > v_c$ and $\rho(0) < \rho_c$. In this case any 
initial condition cannot correspond to an equilibrium state. 
Accordingly, hysteretic effects determine a growth of $\rho(t)$ from
$\rho(0)$, until a value close to $\rho_c$ is approached. Then,
$\rho(t)$ starts to decrease until a linear--in--time decay is
eventually approached, as predicted by Eq.(\ref{rhot}) (as an
example, see Fig.\ref{fig5})~.

The phenomenon of gelation is recovered also in this case, as
predicted on the basis of the order parameter equation. Nonetheless,
a quantitative analysis is prevented for the above mentioned reasons.

\vspace{1mm}
\noindent
{\em 2. The case} $\alpha = 1.5$ 

\begin{figure}[h]
\includegraphics[clip,width=6.5cm]{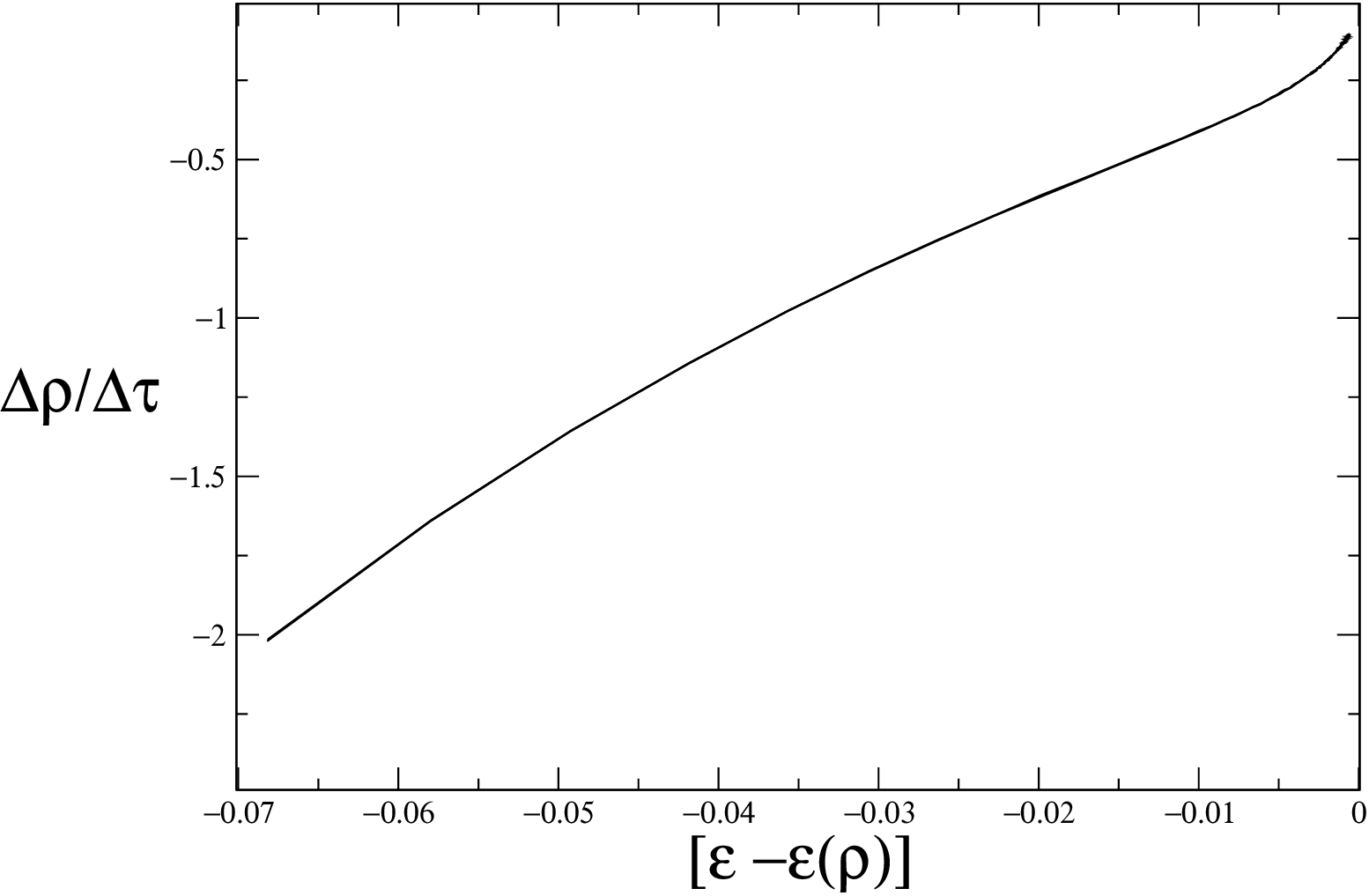}
\caption{
\small
$\dot\rho \equiv \Delta\rho/\Delta\tau$ versus $\epsilon -
\epsilon(\rho)$ for $\alpha =1.5$ and $v = 3.59,\, 3.60,\, 3.61$.
A linear dependence can be attributed to the first few units
of $\tau$.
}
\label{fig6}
\end{figure}

Numerical analysis predicts a critical value of the control parameter
$v_c = 3.60 \pm 0.02$. This result agrees quite well with the theoretical
expectation $v_c = 3.6060508...$. 
On the other hand, the
crossover between the transient and the asymptotic dynamics
is found to extend much more than in the case $\alpha = 2.5$.
For instance, even very close to $v_c$ the transient behavior
ruled by the law $\dot\rho = -\gamma [\epsilon_\rho - \epsilon ]$ 
lasts over a few units of $\tau$ (see Fig.\ref{fig6}). Such a linear
region reduces the more $v$ is far from $v_c$

\begin{figure}[h]
\includegraphics[clip,width=6.5cm]{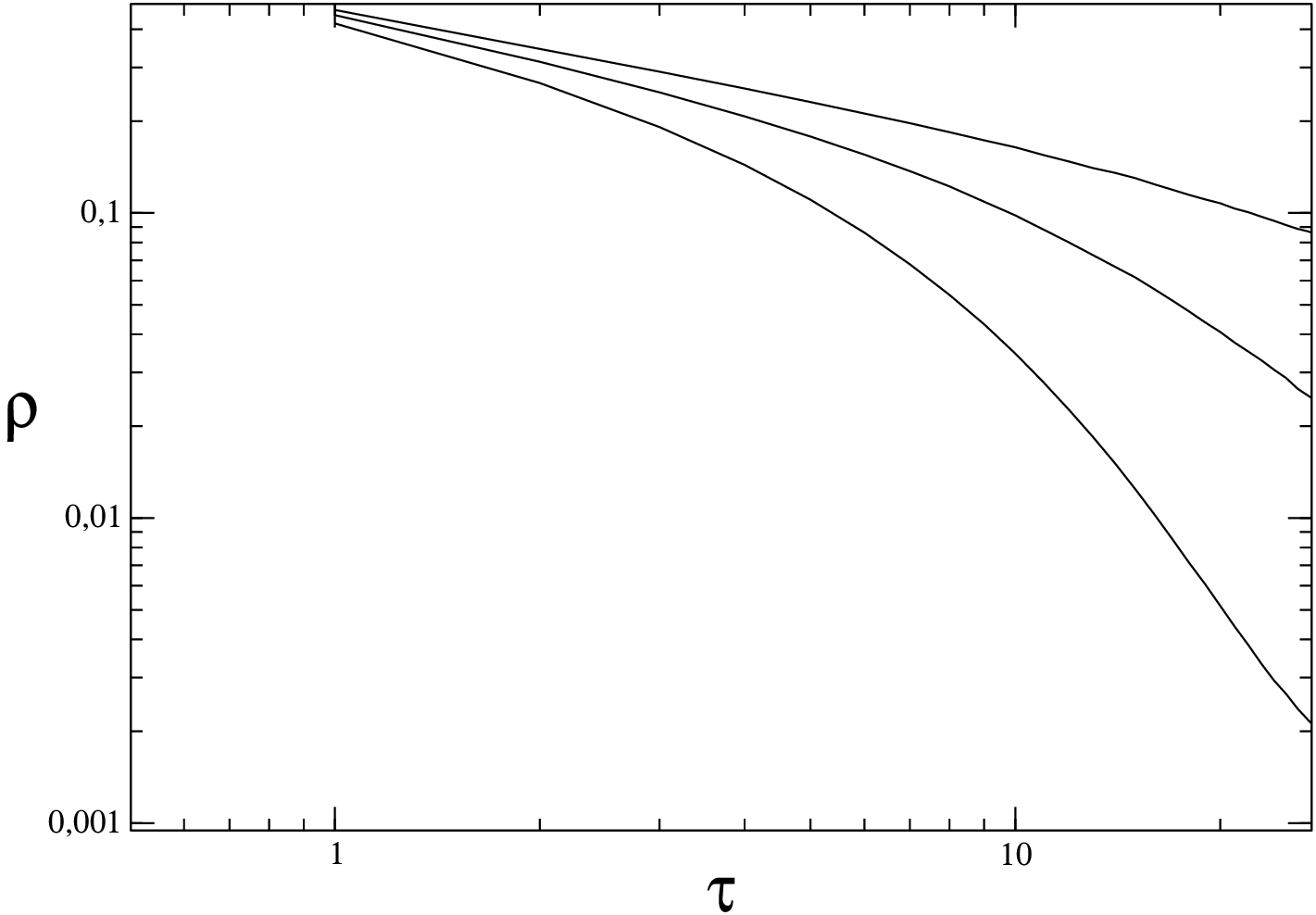}
\caption{
\small
The density of H-bonds $\rho$ as a function of $\tau$ for 
$\alpha =1.5$ and $v = 3.8,\, 4.5,\, 6.0$ (from top to bottom)~.
The double--logarithmic scale shows an aymptotic decay in time
which is faster than linear.
}
\label{fig7}
\end{figure}

However, the long--time dynamics allows to identify the occurrence
of the gelation phenomenon for $v > v_c$ ($T > T_c$). In particular,
$\rho$ eventually exhibits a decay in time which is faster
than linear. In order to exemplify this behavior
in Fig.\ref{fig7} we show $\rho(\tau)$ for some values of $v$ much 
larger than $v_c$.
The asymptotic behavior is ruled by a decay faster than a power-law,
but also slower than an exponential. 
This finding confirms the qualitative agreement with the predicitons 
of the order parameter dynamics.

We want to point out that we have reported the results for 
large values for $v$, because this choice allows to reduce the duration of
the crossover from the transient regime. In fact, for the same values
of $v$ reported in Fig.\ref{fig6} the long--time behavior sets in for
much larger values of $\tau$. 
Moreover, note that also in these cases finite 
size effects do not allow to reach a completely denaturated state
($\rho = 0$). In fact the long time dynamics exhibits fluctuations 
around a low-density regime which is not shown in Fig.\ref{fig7}. 

Also for $v < v_c$ (i.e. $T < T_c$) the transient dynamics 
$\dot\rho = -\gamma [\epsilon_\rho - \epsilon ]$ lasts for
very short times and crosses over to a
slow relaxation to the asymptotic non-zero equilibrium value of
the density, $\rho_{\epsilon} $. As for the case $\alpha =2.5$,
a quantitative study concerning the scaling analysis associated with
the exponential decay rate cannot be worked out satisfactorily with 
our computational resources.

Despite all the difficulties inherent MC simulations, we can conclude that
it exhibits a remarkable qualitative agreement with 
the order parameter dynamics discussed in this Section.

\section{Summary}

In this paper we have derived new results for the classical
Lifson-Poland-Sheraga model of DNA denaturation. In the first part we
have reviewed the main equilibrium properties and completed the list
of known results by the computation of one- and two-point correlation
functions and structure factors, including scaling formulas for the
latter. In the second part we have investigated the dynamical properties
of this model, both numerically and analytically. The Langevin equation
written for the density of H-bonds has been solved with and without the
noise term, in the second case in a linear approximation. Scaling laws for
different values of $\alpha$ have been obtained, and a gelation phenomenon
| arrival to equilibrium in a finite time | above the critical temperature
has been identified. Remarkably, in the cases when comparison
was possible, we have found that the predictions of the  phenomenological
theory were practically indistinguishable from the numerical findings.

Our results suggest that the LPS model belongs to a new universality
class, in the language of critical phenomena, so that most of its details
do not matter near the critical point. This is confirmed by both
the static structure factor and the dynamics of the order parameter.
Concerning the dynamics, the surprising success of the equation of the
order parameter compared with the more detailed description by the
master equation has its counterpart in usual treatments of critical
phenomena. A theoretical explanation of this universality might be given
by a renormalization group analysis of models of the helix-coil transition
that are more microscopic than the LPS one.
However, the best test of universality would be
experimental. We hope that our work will stimulate such experiments,
so that the very nature of the DNA  denaturation transition could finally
be clarified.
A better understanding of the gelation
phenomenon characterizing the high-temperature phase is also missing.
Is this phenomenon basically related to the long--range nature
of the effective interaction of the LPS model? To
our knowledge no systematic investigation has been
performed for identifying such a phenomenon in biomolecules.

Note added: After finishing the analytic part of our work two papers \cite
{BKM}, \cite{SK} have appeared on the preprint archive, dealing with the 
dynamics of the helix-coil transition. Our approach is, however, sensibly 
different.

{\bf Acknowledgments} | We would like to express our gratitude
to the Departments of Physics of the Ecole Polytechnique F\'ed\'erale
de Lausanne and the University of Florence, where the major part of this
work was done. We specially thank Prof. Harald Posch for his kind
hospitality at the Institute of Experimental Physics of the University
of Vienna, where our paper was completed.
The work of A.S. was supported by OTKA Grants T~043494
and T~046129. R.L. wants to thank the Italian Ministry of University
and Research (MIUR), which supported his work through the FIRB project
RBAU01BZJX .


\begin{thebibliography}{99}
\bibitem{L}
S. Lifson, J. Chem. Phys. 40, 3705 (1964)
\bibitem{PS}
D. Poland and H.A. Sheraga,J. Chem. Phys. 45, 1456 (1966)
\bibitem{Causo} M.S. Causo, B. Coluzzi, P. Grassberger, Phys. Rev.
E 62, 3958 (2000).
\bibitem{Carlon} E. Carlon, E. Orlandini and A.L. Stella, Phys. Rev.
Lett. 88, 198101 (2002).
\bibitem{Baiesi} M.Baiesi, E. Carlon and A.L. Stella, Phys. Rev. E 66,
021804 (2002).
\bibitem{BaKa} M.Baiesi, E. Carlon, Y. Kafri, D. Mukamel, E. Orlandini 
and A.L. Stella, Phys. Rev. E 67, 021911 (2003).
\bibitem{Schafer} L. Sch\"afer, cond-mat 0502668 (2005)
\bibitem{Peliti} Y. Kafri, D. Mukamel and L. Peliti, Phys. Rev. Lett.
85, 4988 (2000)
\bibitem{Blossey} R. Blossey and E. Carlon, Phys. Rev. E 68, 061911 (2003).
\bibitem{Wartell} R.M. Wartell and A.S. Benight, Phys. Rep. 126, 67 (1985).
\bibitem{Blake} R.D. Blake and S.G. Delcourt, Nucleic Acids Res. 26, 3323
(1998).
\bibitem{Zocchi} G: Zocchi et al., cond-mat 0304567 (2003).
\bibitem{Bishop2} M. Peyrard and A.R. Bishop, Phys. Rev. Lett. 62, 2755
(1989).
\bibitem{Theodora} N. Theodorakopoulos, T. Dauxois and M. Peyrard, 
Phys. Rev. Lett. 85, 6 (2000).
\bibitem{Nelson} D.R. Nelson, cond-mat 0309559 (2003).
\bibitem{ziff} R.M. Ziff, E.M. Hendriks and M.H. Ernst, Phys. Rev.
Lett. 49, 593 (1982).
\bibitem{BKM}
A. Bar, Y. Kafri and D. Mukamel, cond-mat/0608663
\bibitem{SK}
A. Santos and W. Klein, cond-mat/0610484
\end{thebibliography}
\end{document}